\documentclass[%
 amsmath,amssymb,
 aps, 
prfluids,
]{revtex4-2}

\usepackage{graphicx}% Include figure files
\usepackage{dcolumn}% Align table columns on decimal point
\usepackage{bm}% bold math
\usepackage{xcolor}
\begin{document}

%\title{Interfacial energy transfer beyond capillarity in bubble-laden turbulence} 
% 
\title{Beyond capillarity: extended interfacial energy transfer in bubble-laden turbulence}

\author{Andrea Montessori}
 \affiliation{Department of Civil, Computer Science and Aeronautical Technologies Engineering,
Roma Tre University, Via Vito Volterra, Rome, 00146, Italy}
\email{andrea.montessori@uniroma3.it}

\date{\today}% It is always \today, today,

\begin{abstract}
Capillary forces are commonly regarded as an energetically closed channel in multiphase turbulence, redistributing kinetic energy across scales without net gain or loss. Here we show that this picture breaks down in bubble-laden turbulence when short-range near-contact interactions are active. A spectral analysis of the kinetic-energy budget reveals that the cumulative capillary power does not vanish at statistical stationarity, but instead retains a finite negative residual. This apparent imbalance is exactly compensated by the work of short-range repulsive forces, which is negligible at large and intermediate scales and becomes positive only at small scales. Near-contact interactions therefore open a distinct small-scale pathway in the interfacial energy transfer, rather than acting as a mere regularization of unresolved thin films. The proper scale-by-scale description is thus an extended interfacial transfer, combining capillary and near-contact work, rather than a purely capillary one.
%Capillary forces are widely assumed to be energetically conservative in multiphase turbulence, redistributing kinetic energy across scales without net production or dissipation. Here we show that this paradigm breaks down in bubble-laden turbulence when short-range near-contact interactions are present. A spectral analysis of the kinetic-energy budget reveals that the cumulative capillary power does not vanish at statistical stationarity, but instead exhibits a finite negative residual. This apparent imbalance is exactly compensated by the work of short-range repulsive forces, which is negligible at large and intermediate scales and becomes positive at small scales. These findings demonstrate that, in the presence of  thin-film repulsive interactions, capillarity alone does not provide a closed description of interfacial energy transfer. A consistent energy budget must explicitly account for the work of near-contact interactions, which introduce a distinct small-scale channel restoring global energetic balance.
\end{abstract}

\maketitle

%\tableofcontents

%\section*{}
\paragraph{Introduction.}
Bubble-laden turbulence arises from a continuous competition among inertial transfer,
viscous dissipation, and interfacial dynamics
\cite{mathai2020bubbly,montessori2026breakdown,elghobashi2019direct}. In these flows,
deformable interfaces do not merely respond to turbulence; they actively redirect it
through deformation, breakup, coalescence, and capillary relaxation. As a result, the
kinetic-energy spectrum can depart markedly from the classical Kolmogorov picture and
develop steeper regimes
\cite{mercado2010bubble,almeras2017experimental}, often in connection with a crossover
scale associated with the Hinze mechanism
\cite{crialesi2023interaction,riviere2021sub}. Understanding how interfaces redistribute
kinetic energy across scales is therefore central to the physics of bubbly flows, with
implications for pseudo-turbulence, multiphase mixing, and environmental and industrial
transport processes
\cite{veron2015ocean,angulo2020influence,chandran2015study,bureiko2015current}.

Most scale-by-scale analyses of multiphase turbulence implicitly assume that capillary
work fully exhausts the interfacial contribution to the kinetic-energy budget
\cite{dodd2016interaction}. This assumption was also adopted in our previous work  on bubble-laden HIT \cite{montessori2026breakdown}, where short-range near-contact repulsive forces were not included and the interfacial contribution was therefore identified with capillary work alone. The present Letter addresses a different question: whether this capillary-only description remains energetically closed once near-contact interactions are explicitly active. Such a description is appropriate as long as capillary forces are the only relevant interfacial mechanism. In dense bubbly suspensions, however, interfaces
repeatedly approach each other to distances comparable with the numerical interface
thickness, so that unresolved thin-film dynamics and lubrication-mediated repulsion can
no longer be neglected
\cite{montessori2019mesoscale}.
In this regime, short-range near-contact interaction (NCI)\cite{montessori2019mesoscale,montessori2026adaptive,zhang2022short,RAJIM2026105495} models are introduced to
prevent spurious coalescence and represent the repulsive effect in unresolved films.
Yet, while such models are increasingly used in high-fidelity multiphase simulations,
their energetic role has remained essentially unexplored. Once near-contact forces are
present, it is no longer evident that capillarity alone defines a closed interfacial
channel in the turbulent kinetic-energy budget.

Here we show that it does not. By means of direct numerical simulations of bubble-laden
homogeneous isotropic turbulence, we demonstrate that near-contact interactions break the
classical capillary-guided energy pathway by opening an additional small-scale route in
the spectral kinetic-energy budget. In the absence of NCI, capillary work behaves as an
energetically closed mechanism: it extracts kinetic energy from large scales, reinjects
it at smaller scales, and yields zero net cumulative transfer at statistical stationarity
\cite{dodd2016interaction}. Once near-contact interactions are activated, this closure is
lost at the capillary level alone. Energetic consistency is recovered only by introducing
an \emph{extended interfacial transfer}, defined as the sum of capillary and near-contact
work. We further show that near-contact interactions leave a clear spectral footprint: they sustain kinetic-energy activity beyond the Hinze scale and shift the viscous roll-off to higher wavenumbers. This signature becomes increasingly visible as the Reynolds number decreases and the gas volume fraction increases, revealing a progressive shift from a turbulence-dominated to an interface-dominated regime.

\paragraph{Numerical setup.}
We perform direct numerical simulations of bubble-laden homogeneous isotropic turbulence
using the \texttt{accLB} solver \cite{lauricella2025acclb,lauricella2025thread} on the
Leonardo supercomputing facility. The flow evolves in a triply periodic cubic domain
discretized on a $512^3$ lattice. Turbulence is sustained in a statistically stationary
state by an Arnold--Beltrami--Childress (ABC) forcing \cite{singh2024comparison,dombre1986chaotic} acting
at the largest scales. The initial condition consists of a random suspension of spherical
bubbles with diameter $D \simeq 30$ lattice units. Unless otherwise stated, the
reference case corresponds to a gas volume fraction $\alpha=6\%$. 
In the statistically stationary regime, the characteristic bubble diameter remains close to this value, with $\langle D_b\rangle\simeq30$ lattice units estimated from the curvature distribution. This is very close to the Hinze length, $L_H\simeq34$ lattice units in the reference case, and gives a turbulent Weber number of order unity. The simulated bubbles therefore lie in a deformable, capillary-relevant regime close to the capillary--inertial crossover scale.

The liquid--gas density and viscosity ratios are fixed to
$\rho_l/\rho_g=10^3$ and $\mu_l/\mu_g=10$, respectively, representative of a water--air-like
system, while the surface tension is kept constant at $\sigma=0.01$ in simulation units.
Different Taylor-scale Reynolds numbers are obtained by varying the liquid viscosity while
keeping all other parameters fixed. The case at $Re_\lambda\simeq138$ is used as the main reference because it provides a well-resolved turbulent state in which the turbulent cascade, capillary transfer, and near-contact contribution coexist within the accessible spectral range.  In addition to the reference suspension at $\alpha=6\%$,
we also consider a denser case at $\alpha=12\%$ in order to probe the effect of more
frequent short-range bubble--bubble interactions. Near-contact forces are modeled through
the adaptive repulsive formulation introduced in Ref.~\cite{montessori2026adaptive}. The corresponding near-contact force density can be written in
compact form as
\begin{equation}
\mathbf{f}_{\mathrm{nci}}
=
\rho A_{\mathrm{rep}}\,
\chi_{\mathrm{nc}}\,
q_{\mathrm{pair}}\,
w_h(h)\,
f_{\mathrm{face}}\,
\mathbf{n}_{\mathrm{sym}},
\end{equation}
where $\chi_{\mathrm{nc}}$ activates the interaction only for locally facing
interfaces, $q_{\mathrm{pair}}$ localizes it to the interacting diffuse-interface
pair, $f_{\mathrm{face}}$ measures the opposition of the interface normals, and
$\mathbf{n}_{\mathrm{sym}}$ defines the symmetric film-normal direction.
The factor $w_h(h)=[1+(h/h_0)^p]^{-1}$ localizes the interaction according to
an analytically estimated film-thickness indicator $h$; here $h_0=1$ lattice
unit and $p=4$. In the present work, the associated force is retained explicitly
in the kinetic-energy budget through $W_{\mathrm{nci}}$ and
$S_{\mathrm{nci}}(k)$. 
The present configuration is buoyancy-free: gravity is not included, the domain is triply periodic, and kinetic energy is injected externally by the large-scale ABC forcing. Therefore the bubbles do not possess a mean buoyancy-driven rise velocity, and quantities such as the Galileo number, Bond number, or $u'/V_b$ are not controlling parameters of the present problem. The relevant dimensionless groups are instead $Re_\lambda$, the gas volume fraction $\alpha$, the density and viscosity ratios, and the turbulent Weber number based on the bubble diameter, $We_D=\rho_l u_{\mathrm{rms}}^2D/\sigma$,as further discussed in the Supplemental Material \cite{SupplementalMaterial}. This distinguishes the present configuration from buoyancy-driven or rising-bubble systems, where gravity, bubble rise, and wake production provide the dominant energy-injection mechanism \cite{huang2025taylor,mathai2025swarming}.

\paragraph{Spectral energy budget.} To analyze how interfacial forces redistribute kinetic energy across scales, we frame the results in terms of an extended turbulent kinetic-energy budget. At the
domain-averaged level,
\begin{equation}
\frac{\mathrm{d}\mathcal{K}}{\mathrm{d}t}
=
\mathcal{P}_{\mathrm{ABC}}
-\varepsilon
+\mathcal{W}_{\sigma}
+\mathcal{W}_{\mathrm{nci}},
\label{eq:tke_global}
\end{equation}
where $\mathcal{K}=\langle |\mathbf{u}|^2/2\rangle$ is the mean kinetic energy,
$\mathcal{P}_{\mathrm{ABC}}=\langle \mathbf{u}\cdot\mathbf{f}_{\mathrm{ABC}}\rangle$ is
the power injected by the ABC forcing, $\varepsilon$ is the mean viscous dissipation
rate, and $\mathcal{W}_{\sigma}=\langle \mathbf{u}\cdot\mathbf{f}_{\sigma}\rangle$ and
$\mathcal{W}_{\mathrm{nci}}=\langle \mathbf{u}\cdot\mathbf{f}_{\mathrm{nci}}\rangle$ are
the mean capillary and near-contact work, respectively. In spectral space, the
shell-integrated budget reads
\begin{equation}
\partial_t E(k)
=
T(k)+F_{\mathrm{ABC}}(k)-D(k)+S_\sigma(k)+S_{\mathrm{nci}}(k),
\label{eq:tke_spectral}
\end{equation}
where $T(k)$ is the nonlinear transfer, $F_{\mathrm{ABC}}(k)$ the forcing input, and
$D(k)$ the viscous dissipation spectrum.

The interfacial contributions $S_\sigma(k)$ and $S_{\mathrm{nci}}(k)$ are computed as
shell-integrated velocity--force co-spectra,
\begin{equation}
S_\sigma(k)=
\sum_{k-\Delta k/2<|\mathbf{q}|\le k+\Delta k/2}
\left\langle
\Re\!\left[
\hat{\mathbf{u}}(\mathbf{q},t)\cdot
\hat{\mathbf{f}}_{\sigma}^{\,*}(\mathbf{q},t)
\right]
\right\rangle,
\qquad
%\end{equation}
%\begin{equation}
S_{\mathrm{nci}}(k)=
\sum_{k-\Delta k/2<|\mathbf{q}|\le k+\Delta k/2}
\left\langle
\Re\!\left[
\hat{\mathbf{u}}(\mathbf{q},t)\cdot
\hat{\mathbf{f}}_{\mathrm{nci}}^{\,*}(\mathbf{q},t)
\right]
\right\rangle,
\end{equation}
where hats denote Fourier transforms, ${}^*$ indicates complex conjugation, and angular
brackets denote averaging in the statistically stationary regime. The corresponding
cumulative transfers are
\begin{equation}
\Pi_\sigma(k)=\int_0^k S_\sigma(k')\,\mathrm{d}k',
\qquad
\Pi_{\mathrm{nci}}(k)=\int_0^k S_{\mathrm{nci}}(k')\,\mathrm{d}k'.
\end{equation}
When near-contact interactions are active, we define the extended interfacial transfer
\begin{equation}
S_{\mathrm{ext}}(k)=S_\sigma(k)+S_{\mathrm{nci}}(k),
\qquad
\Pi_{\mathrm{ext}}(k)=\Pi_\sigma(k)+\Pi_{\mathrm{nci}}(k).
\end{equation}
Under the statistically stationary conditions considered here, energetic closure of the
interfacial channel requires $\Pi_{\mathrm{ext}}(k)\to 0$ at sufficiently large $k$. As
shown below, this condition is satisfied only by the combined capillary--near-contact
contribution: once near-contact interactions are active, the capillary term alone is no
longer closed.

To identify the crossover between turbulence-dominated and interface-dominated scales, we estimate the Hinze
length \cite{deane2002scale} as
$L_H = C_H (\sigma/\rho_l)^{3/5}\epsilon_{\mathrm{avg}}^{-2/5}$, where $\epsilon_{\mathrm{avg}}$ is the mean kinetic-energy
dissipation rate. In the present work we set $C_H=0.78$, following Deane and Stokes \cite{deane2002scale}. The corresponding Hinze wavenumber, $k_H = 2\pi/L_H$, is used as an organizing scale for the crossover between turbulence-dominated and interface-dominated dynamics, rather than as a sharply defined transition wavenumber. Since $C_H$ is an order-one empirical constant, moderate variations of its value would shift the marker for $k_H$ by an order-one factor but would not alter the qualitative interpretation of the spectral reorganization around and beyond the Hinze scale.

\begin{figure}
    \centering
    \includegraphics[width=0.9\linewidth]{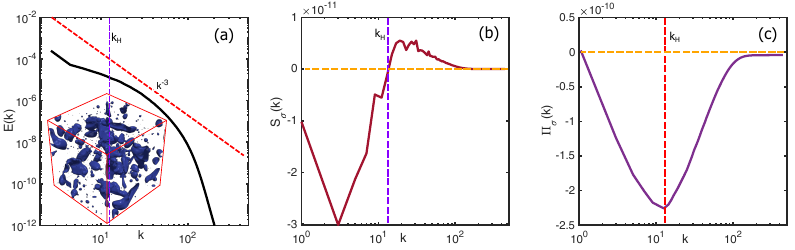}
    \caption{Baseline capillary energy transfer in bubble-laden homogeneous isotropic
    turbulence at gas volume fraction $\alpha=6\%$ and $Re_\lambda=138$, without
    near-contact interactions. (a) Turbulent kinetic-energy spectrum $E(k)$, showing a finite resolved range
consistent with an effective slope close to the $k^{-3}$ reference behavior. (b) Capillary power spectrum
    $S_\sigma(k)$, which is negative at low wavenumbers, crosses zero around the Hinze
    wavenumber $k_H$, and becomes positive at large $k$. Capillary forces therefore
    extract kinetic energy from large scales and re-inject it at smaller ones. (c)
    Cumulative capillary transfer $\Pi_\sigma(k)$, which first decreases, reaches a
    minimum, and then returns to zero at large $k$, as expected for a closed interfacial
    redistribution pathway at statistical stationarity.}
    \label{fig:no_nci_spectra}
\end{figure}

\paragraph{Results.}
We first establish the classical interfacial picture in the absence of near-contact
interactions. For the reference bubble-laden case at $\alpha=6\%$ and
$Re_\lambda \simeq 138$, Fig.~\ref{fig:no_nci_spectra}(a) shows a turbulent
kinetic-energy spectrum with a finite resolved range consistent with an effective
slope close to the $k^{-3}$ reference behavior. 
In this baseline configuration, capillarity acts as a closed redistribution mechanism across
scales. As shown in Fig.~\ref{fig:no_nci_spectra}(b), the capillary work spectrum
$S_\sigma(k)$ is negative at low wavenumbers, crosses zero around the Hinze wavenumber
$k_H$, and becomes positive at larger $k$. Capillary forces therefore extract kinetic
energy from the larger, bubble-deforming scales and return it at smaller scales, where
interfacial relaxation becomes dominant. Accordingly, the cumulative transfer
$\Pi_\sigma(k)$ first decreases, reaches a minimum, and then returns to zero at large
$k$ \cite{pandey2022turbulence}, as shown in Fig.~\ref{fig:no_nci_spectra}(c). Without
NCI, capillarity alone provides a closed interfacial energy pathway.
\begin{figure}
    \centering
    \includegraphics[width=0.55\linewidth]{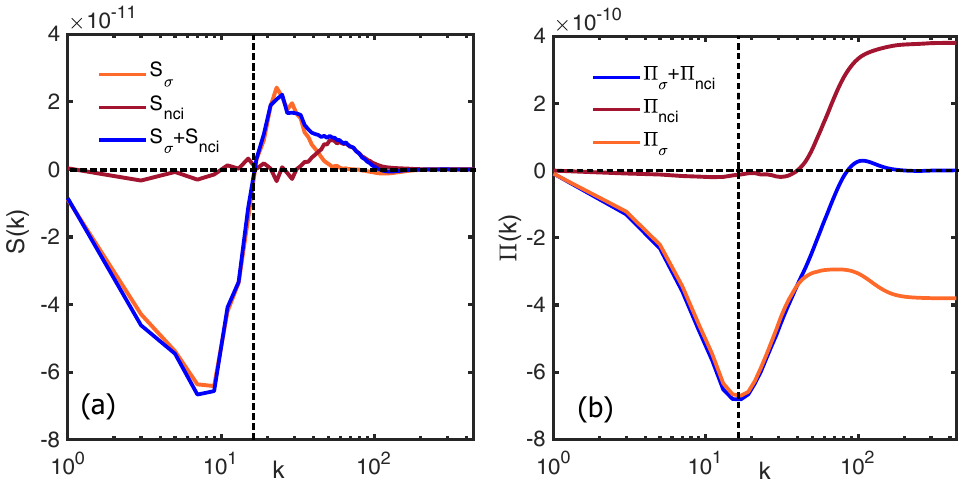}
    \caption{Breakdown of capillary closure in the reference bubble-laden HIT case at
    gas volume fraction $\alpha=6\%$ and $Re_\lambda=138$ when near-contact interactions
    are included. (a) Power spectra of capillary work $S_\sigma(k)$, near-contact work
    $S_{\mathrm{nci}}(k)$, and their sum. The near-contact contribution is negligible at
    low and intermediate wavenumbers and becomes positive only at sufficiently high $k$,
    thereby opening a distinct small-scale channel of interfacial work. (b)
    Corresponding cumulative transfers. The cumulative capillary work
    $\Pi_\sigma(k)$ retains a finite negative residue at large $k$, whereas
    $\Pi_{\mathrm{nci}}(k)$ provides the compensating positive transfer required for the
    total cumulative interfacial transfer
    $\Pi_{\mathrm{ext}}(k)=\Pi_\sigma(k)+\Pi_{\mathrm{nci}}(k)$ to vanish.}
    \label{fig:S_pi_plots}
\end{figure}
The central result of this work is that this capillary closure is broken once
near-contact interactions are activated. Figure~\ref{fig:S_pi_plots} shows that, for the
same reference case, $S_{\mathrm{nci}}(k)$ is essentially negligible at low and
intermediate wavenumbers and becomes positive only at high $k$. Near-contact forces
therefore do not alter the large-scale dynamics directly; instead, they open a distinct
small-scale channel of positive interfacial work. The effect on the cumulative budget is
decisive: $\Pi_\sigma(k)$ no longer returns to zero, but saturates to a finite negative
residual at large wavenumbers. 
\begin{figure}
    \centering
    \includegraphics[width=0.32\linewidth]{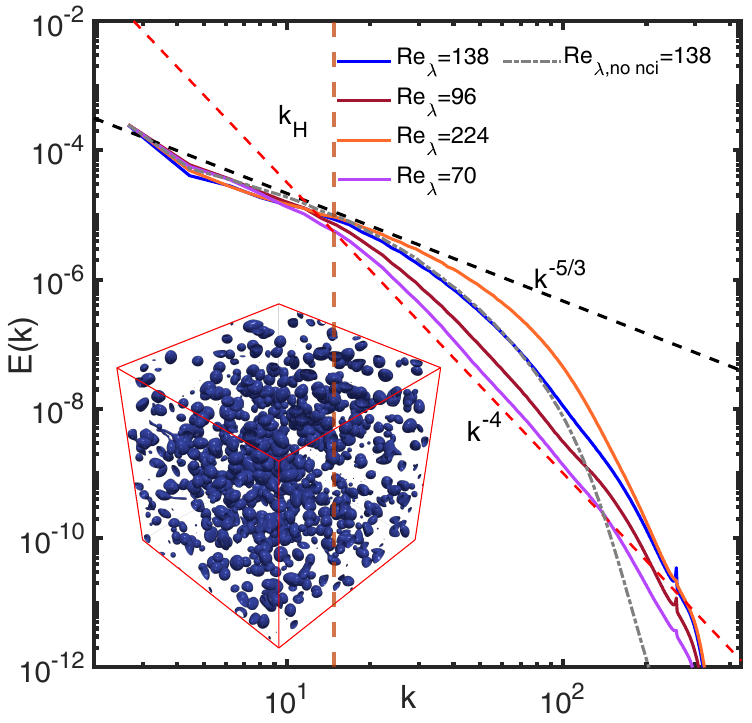}
    \caption{Turbulent kinetic-energy spectra for the bubble-laden HIT configuration at
    gas volume fraction $\alpha=6\%$, shown for decreasing Taylor-scale Reynolds number
    $Re_\lambda$, in the presence of near-contact interactions (solid lines). At the highest Reynolds number, $Re_\lambda=224$, the spectrum exhibits an extended range consistent with an effective slope close to the Kolmogorov $k^{-5/3}$ reference scaling. As $Re_\lambda$ is reduced to 138, 96, and 70, the spectra
progressively steepen, with effective slopes becoming comparable to $k^{-3}$ and then approaching a steeper $k^{-4}$ reference behavior over the plotted range. The crossover
    occurs around the Hinze wavenumber $k_H$, indicating that interfacial dynamics
    increasingly reshape the spectrum beyond the Hinze scale. The dashed curve shows the corresponding case at $Re_\lambda=138$ without NCI, highlighting that near-contact interactions shift the viscous roll-off to higher wavenumbers and extend the active spectral range. The power-law lines are intended as visual reference slopes over finite resolved ranges.}
    \label{fig:tke_comp}
\end{figure}
Capillarity alone therefore ceases to define a closed
interfacial redistribution channel. Energetic closure is recovered only when the
near-contact contribution is included, since $\Pi_{\mathrm{nci}}(k)$ supplies the
positive compensation required for
$\Pi_{\mathrm{ext}}(k)=\Pi_\sigma(k)+\Pi_{\mathrm{nci}}(k)$ to vanish. The relevant
quantity entering the scale-by-scale kinetic-energy budget is therefore not the capillary
transfer alone, but the extended interfacial transfer.

The robustness of this result is assessed in the Supplemental Material \cite{SupplementalMaterial} by varying both the near-contact repulsive amplitude and the diffuse-interface thickness. These tests show that, once spurious coalescence is avoided and the interface remains resolved, the cumulative NCI contribution remains positive at small scales and preserves its compensating role in the extended interfacial budget.

A direct closure check of the full shell-integrated kinetic-energy budget is also reported in the Supplemental Material \cite{SupplementalMaterial}. The residual remains within the numerical and statistical uncertainty of the simulations, confirming that the apparent imbalance of the capillary contribution is removed only when the near-contact work is included.

Having established that near-contact interactions open an additional small-scale
interfacial pathway, we now examine how this affects the shape of the kinetic-energy
spectrum. Figure~\ref{fig:tke_comp} summarizes the Reynolds-number dependence at fixed
gas volume fraction, $\alpha=6\%$, with $Re_\lambda$ varied from about 224 down to 70.
At the highest Reynolds number, the spectrum retains an extended range consistent with an effective slope close to the
Kolmogorov $k^{-5/3}$ reference scaling over the resolved range. As $Re_\lambda$ is reduced to 138, 96, and 70, the
spectrum progressively steepens, first toward an effective slope comparable to $k^{-3}$ and then toward a steeper range approaching $k^{-4}$. These power laws are used here as reference slopes over finite resolved intervals, rather than as evidence of asymptotic scaling regimes. In particular, the $k^{-4}$ line is not intended as a universal scaling law, but only as a visual reference for the effective steepening observed in the lowest-$Re_\lambda$ cases over a finite sub-Hinze range, where nonlinear transfer becomes comparatively weaker and the extended interfacial contribution is increasingly balanced by viscous dissipation. The crossover occurs around $k_H$, indicating that the Hinze scale
marks the transition from a turbulence-dominated range to one increasingly controlled by
interfacial dynamics. At the same time, the comparison at $Re_\lambda \simeq 138$ between the cases with and without NCI shows that near-contact interactions shift the final viscous roll-off to higher wavenumbers, keeping the spectrum active over a broader range of scales. 
The no-NCI simulation at the same gas volume fraction and forcing protocol provides the direct baseline for isolating the energetic effect of near-contact interactions. Single-phase HIT and capillary-only bubble-laden reference cases were analyzed in detail in Ref.~\cite{montessori2026breakdown}.
This behavior is consistent with an extended pseudo-inertial range sustained by the additional small-scale interfacial transfer.
We also note that decreasing $Re_\lambda$ weakens the classical inertial cascade and makes the action of the extended interfacial transfer progressively more visible, consistent with the emergence of a spectral regime increasingly controlled by interfacial work and viscous dissipation. This trend is qualitatively reminiscent of previous observations in bubbly
pseudo-turbulence \cite{pandey2023kolmogorov}.
\begin{figure}
    \centering
    \includegraphics[width=0.65\linewidth]{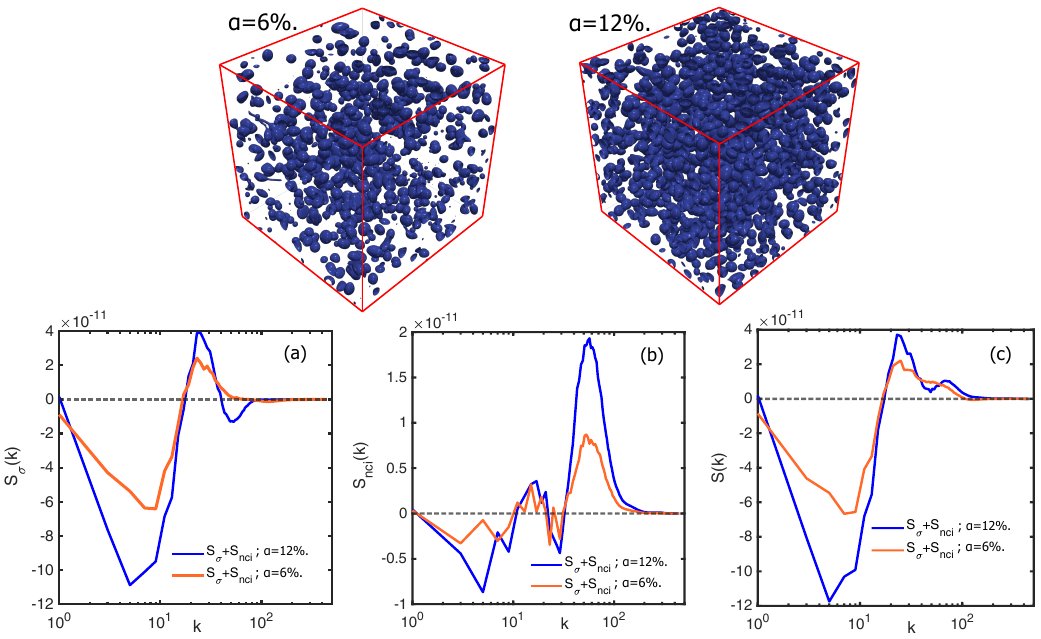}
    \caption{Effect of increasing gas volume fraction on the extended interfacial
    transfer at fixed $Re_\lambda=138$. Upper panel: instantaneous bubble
    configurations for $\alpha=6\%$ and $\alpha=12\%$. (a) Near-contact power spectra,
    showing that increasing bubble concentration amplifies the short-range contribution
    in the same spectral region where capillary effects are strongest. (b) Interfacial
    spectra, showing that enhanced near-contact activity induces a readjustment of the
    capillary work: $S_\sigma(k)$ develops a secondary negative lobe at higher
    wavenumbers before relaxing to zero. In the total interfacial transfer (c), this
    readjustment appears as a weaker secondary positive peak at high $k$, highlighting
    the increasingly important role of near-contact interactions in dense suspensions.}
    \label{fig:S_comp_nbbls}
\end{figure}
A similar amplification is observed when the gas volume fraction is increased at fixed
$Re_\lambda \simeq 138$. As shown in Fig.~\ref{fig:S_comp_nbbls}, raising the bubble
concentration from $\alpha=6\%$ to $\alpha=12\%$ markedly strengthens the near-contact
contribution in the same spectral range where interfacial activity is already strongest.
At the same time, the capillary spectrum itself reorganizes:
$S_\sigma(k)$ develops a secondary negative lobe at higher wavenumbers before relaxing to
zero. In the total interfacial spectrum, this readjustment appears as a weaker secondary
positive peak at high $k$. Increasing bubble concentration therefore does not simply
magnify the short-range contribution; it restructures the small-scale interfacial
transfer. The emerging picture is that of a two-stage pathway: a primary capillary
redistribution organized around the Hinze scale, and a secondary NCI-mediated channel at
smaller scales that becomes increasingly important as near-contact events become more
frequent. To connect this spectral signature with the underlying interface dynamics, we also monitored the number of active near-contact events, $N_{\mathrm{act}}$, defined as the number of interface regions satisfying the NCI activation criterion in statistically stationary snapshots. At fixed $Re_\lambda\simeq138$, $N_{\mathrm{act}}$ increases from approximately $7\times10^4$ at $\alpha=6\%$ to approximately $2\times10^5$ at $\alpha=12\%$. Thus, doubling the gas volume fraction produces an approximately three-fold increase in the number of activated near-contact events. This increase is consistent with the amplification of the high-wavenumber near-contact contribution and supports the interpretation that the positive small-scale NCI work originates from repeated localized near-contact interactions between neighboring bubbles.

Taken together, these results show that the classical capillary picture of interfacial
energy transfer becomes incomplete once short-range repulsive interactions are dynamically
relevant. In that regime, kinetic energy is no longer redistributed through capillary
work alone: an additional small-scale pathway emerges and must be included to recover a
closed interfacial budget. The appropriate scale-by-scale description is therefore not a
purely capillary one, but an extended interfacial transfer that combines capillary and
near-contact work.
\paragraph{Conclusions.}
We have shown that, in bubble-laden turbulence, the interfacial contribution to the
spectral kinetic-energy budget ceases to be closed at the capillary level once
short-range near-contact interactions become active. In the absence of such interactions,
capillary work redistributes kinetic energy across scales and yields zero net cumulative
transfer at statistical stationarity. When unresolved thin-film dynamics are represented
through short-range repulsive forces, an additional positive small-scale contribution
appears and provides the compensation required to restore closure of the interfacial
budget. This identifies an extended interfacial transfer, combining capillary and
near-contact work, as the proper quantity entering the scale-by-scale kinetic-energy
balance.

The effect becomes more pronounced as the Reynolds number decreases and the gas volume
fraction increases, indicating a progressive shift toward an interface-dominated regime.
In the denser suspension, this amplification is accompanied by a larger number
of activated near-contact events, providing a direct link between the spectral NCI work
and the underlying interface dynamics.

More broadly, our results show that whenever near-contact physics is dynamically relevant,
the standard capillary-only description of interfacial energy transfer is no longer
sufficient. In that regime, short-range repulsive interactions are not a negligible
correction, but an intrinsic part of the multiscale pathway through which kinetic energy
is redistributed in dense bubbly flows. This result also provides a concrete route for incorporating near-contact physics into reduced-order or subgrid-scale descriptions of dense
multiphase turbulence. When thin-film and near-contact dynamics are not
explicitly resolved, their effect should enter as an additional interfacial-work
closure. The present DNS constrains such a contribution to be negligible at
large and intermediate scales, positive at small scales, and to compensate the
finite capillary residual so that the total interfacial transfer remains
energetically closed. The observed increase of $N_{\mathrm{act}}$ with gas
volume fraction further suggests that local near-contact activity, or an
equivalent measure of interface crowding, may provide a natural state variable
for modulating the strength of this unresolved contribution.

\paragraph*{\textbf{Acknowledgments}}
A.M. acknowledges funding from the Italian Government through the PRIN project MOBIOS (Grant No. 2022N4ZNH3, CUP F53C24001000006) and computational support from CINECA through the ISCRA B project MIPLAST (IsB31, HP10BZY7BK). Fruitful discussions with Sauro Succi are also gratefully acknowledged.

%\nocite{*}

\providecommand{\noopsort}[1]{}\providecommand{\singleletter}[1]{#1}%
\end{document}